\documentclass{SCIS2019}
\usepackage{amsmath}
\usepackage{amssymb}
\usepackage{graphicx}
\usepackage{tipa}
\usepackage{upgreek}
\usepackage{dsfont}

\usepackage{graphicx}
\usepackage{caption}

\begin{document}
\ArticleType{RESEARCH PAPER}
\Year{2019}
\Month{}
\Vol{}
\No{}
\DOI{}
\ArtNo{}
\ReceiveDate{}
\ReviseDate{}
\AcceptDate{}
\OnlineDate{}

\title{Analysis of 3D Localization in Underwater Optical Wireless Networks with Uncertain Anchor Positions}

\author[1]{Nasir SAEED}{mr.nasir.saeed@ieee.org}
\author[1]{Abdulkadir CELIK}{abdulkadir.celik@kaust.edu.sa}
\author[1]{Mohamed-Slim ALOUINI}{}
\author[1]{Tareq Y. AL-NAFFOURI}{}


\AuthorMark{Nasir Saeed}

\AuthorCitation{Nasir Saeed, Abdulkadir Celik, Mohamed-Slim Alouini, Tareq Al-Naffouri. Analysis of 3D Localization in Underwater Optical Wireless Networks with Uncertain Anchor Positions}


\address[1]{Computer, Electrical, Mathematical Sciences \& Engineering (CEMSE) Division, \\
 King Abdullah University of Science and Technology (KAUST),\\ 
 Thuwal, 23955, KSA.}


\abstract{
Localization accuracy is of paramount importance for the proper operation of underwater optical wireless sensor networks (UOWSNs). However, underwater localization is prone to hostile environmental impediments such as drifts due to the surface and deep currents. These cause uncertainty in the deployed anchor node positions and pose daunting challenges to achieve accurate location estimations. Therefore, this paper analyzes the performance of three-dimensional (3D)  localization for UOWSNs and derive a closed-form expression for the Cramer Rao lower bound (CRLB) by using time of arrival (ToA) and angle of arrival (AoA) measurements under the presence of uncertainty in anchor node positions.  Numerical results validate the analytical findings by comparing the localization accuracy in scenarios with and without anchor nodes position uncertainty. Results are also compared with the linear least square (LSS) method and weighted LLS (WLSS) method.
}

\maketitle
\section{Introduction}
Underwater wireless communications (UWCs) can be carried out by different wireless carrier types such as radio frequency (RF) signals, acoustic waves, magnetic induction (MI), and light beams. RF-based UWC is greatly affected by high attenuation because the nature of seawater is conductive, and therefore, it requires huge antennas and power-hungry transceivers. For this reason, acoustic UWC systems are the most commonly used UWC systems due to their long transmission range in the order of kilometers. Nevertheless, acoustic waves have low data rates, limited bandwidth, and high latency because of their low propagation speed \cite{Saeed2017}. Alternatively, underwater optical wireless communication (UOWC) is desirable for their high data rates in the order of Gbps and very low latency thanks to the high propagation speed, energy efficiency, and low-cost \cite{Zeng2017}. Therefore, UOWC has recently been considered as a promising technology for underwater optical wireless sensor networks (UOWSNs) in many scientific, industrial, and military applications \cite{Saeed2018Survey}.

The data gathered by the sensors need to be geographically tagged to provide valuable information. Location is also critical for proper functioning of the pointing and alignment mechanism, which is closely related to the link reliability. Furthermore, accurate localization is necessary to develop routing protocols to mitigate the short-range limitation of UOWCs and enhance end-to-end network performance. UOWSN localization is more demanding than terrestrial wireless networks due to the unavailability of GPS signals and unique propagation characteristics of light in the aquatic medium \cite{Nasir2018limited}. Recently, numerous localization techniques for UOWSNs have been proposed in two-dimensional space \cite{Nasir2018limited, Akhoundi2017underwater, Nasir2018twc, Nasir2018tmc}, however, the directed nature of UOWCs calls for novel 3D localization techniques. In \cite{Nasir2018spawc}, a 3D localization technique has been proposed in the presence of outliers where the position of anchors is assumed to be perfectly known. However, the accuracy of any localization technique depends on the type of ranging and the anchor positions \cite{Mekonnen2014}. The problem of anchor node uncertainty is studied in the past for various indoor localization techniques. For example, an RSS-based localization method was proposed in \cite{kumar2017wls} with perturbed anchor locations where a weighted least square solution was used. Similarly, semi-definite programming was used in \cite{Lui2009} and \cite{suliman2018robust}, for a ToA based wireless sensor network localization with uncertain anchor positions. However, none of the above works consider a joint ToA and AoA-based approach with uncertain anchor position in UOWSNs. In practice, the underwater environment is dynamic in nature, which causes anchors to drift apart from their actual positions. Therefore, it is required to develop a localization technique that considers the anchor node position uncertainty. The major contributions of this paper can be summarized as follows: 
\begin{itemize}
\item Firstly, the Cramer Rao lower bound (CRLB) is derived for 3D UOWSNs localization with the joint ToA and AoA ranging measurements. The CRLB characterizes the lower bound for the error variance of any unbiased estimator \cite{Jia2008}.
\item Secondly, the analysis is extended to the case when there is uncertainty in the anchor positions, which is more practical and challenging. Finally, the results for the CRLB are compared with and without uncertainty in the anchor positions.
\item  Finally, the performance of the LSS \cite{Cheung2004} and WLLS \cite{Tarrio2008} solutions are compared with the CRLB. Numerical results show that the uncertainty in the anchor positions provides a practical bound.
\end{itemize} 

The remainder of the paper is organized as follows: Section II formulates the localization problem with uncertainty in the anchor positions. Section III derives the CRLB for 3D localization with anchor position uncertainty (APU). Following the numerical results in Section IV, Section V concludes the paper.

\section{Problem Formulation}
\begin{figure}[t]
\centering
\captionsetup{justification=centering,font=small}
\includegraphics[width=0.5 \columnwidth]{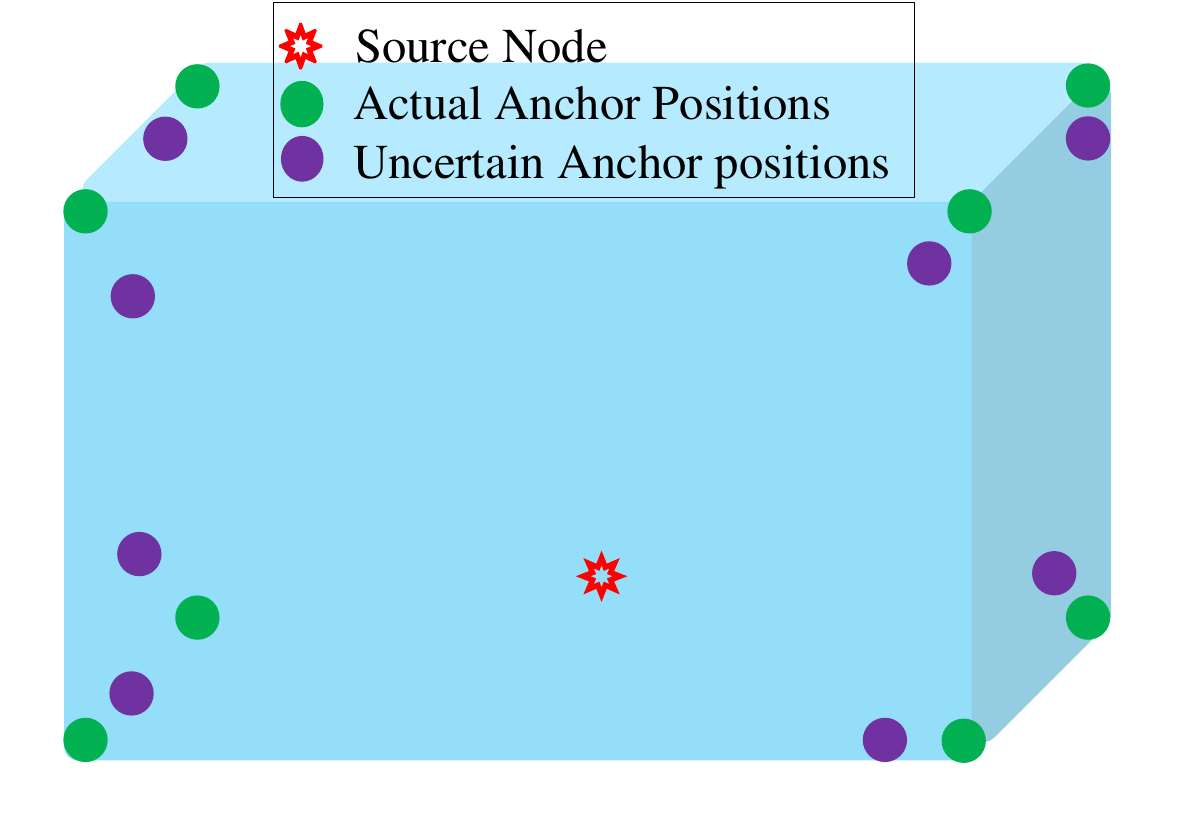}  
\caption{System model.}
\label{fig:model}  
\end{figure}

We consider a 3D-UOWSN where a source node and $M$ anchor nodes are randomly deployed as shown in Fig.~\ref{fig:model}. The location of the source node is unknown, whereas that of anchor nodes is known a priori. The propagation loss of UOWC channel is mainly driven by the absorption and scattering effects of the aquatic medium. Based on the commonly used Beer-Lambert channel model, the range between two nodes can be measured by the received signal strength \cite{Nasir2018limited, Nasir2018twc}. 
In this work, however, the range measurements between the source node and the anchor nodes are estimated by using both ToA and AoA methods. ToA method measures the distance to an anchor node $i$ based on the measured propagation time $t_i$ and underwater speed of light $c$, i.e., $d_i=c\cdot t_i$ where $c \approx 2.55 \times 10^8$. On the other hand, we assume that the source node has multiple photo-diodes in different directions to receive signals from multiple anchor nodes. Based on the signal received at different photo-diodes from different anchors, the azimuth and elevation angles are calculated.

Moreover, we consider that the range measurements are corrupted by Gaussian noise, which is a common assumption used in the literature \cite{van2014, ihsan2019}. Also, a more realistic assumption of the uncertainty in anchor positions is introduced. Likewise, the UOWC channel changes rapidly, and therefore to reduce the effect of its temporal nature, we consider the average value of both the ToA and AoA range measurements.
The actual position of the source node is denoted by $\boldsymbol{\ell} = [x,y,z]^T$ and the uncertain location of $j$-th anchor is denoted by $\hat{\boldsymbol{\ell}}_j = [\hat{x}_j,\hat{y}_j,\hat{z}_j]^T$. The uncertainty in the anchor positions is not exactly known and thus modeled as a random variable following a normal distribution with zero mean and covariance $\boldsymbol{\mathbf{C}}_j$ which also defines the prior distribution of uncertainty in the $j$-th anchor node position. The purpose of the proposed theoretical analysis is to provide a lower bound on the error variance of the 3D localization for a source node given that the noisy range measurements are available between the source node and $j$-th anchor node with the uncertain position. 

In order to evaluate the performance of any localization technique, several metrics are used, such as geometric dilution of precision (GDOP), root mean squared error (RMSE), and circular error probability (CEP). In this paper, we focus on the variance of the location estimation that is defined by the CRLB. It is worth noting that the derived RMSE of the location estimator is equivalent to the CRLB in case of unbiased estimation. In the following, we derive the CRLB for joint ToA and AoA measurements, and then we extend the derivation to the case where there is uncertainty in the anchor positions. Based on \cite{Nasir2018limited, Akhoundi2017underwater}, the noisy range measurements for UOWC can be written as $\text{\textramshorns}_i = g_i(\boldsymbol{\ell})+\eta_i, i \in [1,2,...,3M]$ where the first term,
\begin{equation}
\label{eq:measurement}
g_i(\boldsymbol{\ell})= \begin{cases}
d_j = \sqrt{\tilde{x}_j^2+\tilde{y}_j^2+\tilde{z}_j^2} &, i=3j-2 \\
\phi_j = \tan^{-1}\left(\frac{\tilde{y}_j}{\tilde{x}_j}\right) &, i=3j-1 \\
\theta_j=\cos^{-1}\left(\frac{\tilde{z}_j}{g_{3j-2}(\boldsymbol{\ell})}\right) &,i=3j \\
\end{cases}, 
\end{equation}
represents the ToA and AoA measurements for $i=3j-2$ and $i={3j-1,3j}$, respectively. The terms $\tilde{x}_j$,  $\tilde{y}_j$, and $\tilde{z}_j$ are equal to $x-x_j$, $y-y_j$, and $ z-z_j$, respectively. $\eta_i$ is the measurement noise which is modeled as zero mean Gaussian random variable with variances equal to $\sigma_{\hat{d}_j}^2$, $\sigma^2_{\hat{\phi}_j}$, and $\sigma^2_{\hat{\theta}_j}$ for $i=3j-2$, $i=3j-1$, and $i=3j$, respectively \cite{Patwari2005}. The subscripts $\hat{d}_j$, $\hat{\phi}_j$, and $\hat{\theta}_j$ correspond to the estimated distance, azimuth angle, and elevation angle from the source node to the $j$-th anchor node, respectively. The range measurements can be written in the $1 \times 3M$ vector forms as $ \textbf{\textramshorns} = \boldsymbol{g}_i(\boldsymbol{\ell}) +  \boldsymbol{\eta}$. Since the ranging errors are modeled as Gaussian random variables, the probability density function for range measurements is given by
 \begin{equation}\label{eq: pdf}
 p(\textbf{\textramshorns}|\boldsymbol{\ell})=\frac{1}{\sqrt{(2\pi)^{3M}}\Pi_{j=1}^{3M}\sigma_{n_j}}\exp^{\left(-\sum_{j=1}^{3M}\frac{(\text{\textramshorns}_j-g_{j}(\boldsymbol{\ell}))^2}{2\sigma_{n_j}^2}\right)}.
 \end{equation}
The CRLB is equivalent to the inverse of the Fisher information matrix (FIM), which is given in \cite{Patwari2005} as
\begin{equation}
 \text{CRLB}(\tilde{\boldsymbol{\ell}}_k) = [\boldsymbol{\mathcal{J}}^{-1}(\boldsymbol{\ell})]_{k,k},
\end{equation}
where $\boldsymbol{\mathbf{J}}(\boldsymbol{\ell})$ is the FIM and $k \in \{1,2,3\}$ for 3D localization. The elements of FIM are calculated by taking the partial derivatives of the log likelihood function \eqref{eq: pdf} with respect to each location coordinate as follows
\begin{equation}
[\boldsymbol{\mathbf{J}}(\boldsymbol{\ell})]_{k,q} = -\mathds{E}\left( \frac{\partial^2 \log( p(\textbf{\textramshorns}|\boldsymbol{\ell}))}{\partial\boldsymbol{\ell}_k\partial\boldsymbol{\ell}_q}\right),
\end{equation}
where $\mathds{E}(\cdot)$ represents the expectation operator and $k,~q \in\{1,2,3\}$. After some manipulations, the FIM elements can be written as
 \begin{eqnarray}
 [\boldsymbol{\mathbf{J}}(\boldsymbol{\ell})]_{1,1}&=&  \sum_{j=1}^M 3 k_j \mu \left[ \bigg(\frac{\tilde{x}_j}{\sigma_{d_j}d_j}\biggl)^2+ \bigg(\frac{\tilde{y}_j}{\sigma_{\phi_j}d_{2,j}^2}\biggl)^2 \right.\nonumber\\
 & & \left. + \bigg(\frac{\tilde{x}_j\tilde{z}_j}{\sigma_{\theta_j} d_j^2 d_{2,j}}\bigg)^2 \right],
 \end{eqnarray}
 \begin{eqnarray}
 [\boldsymbol{\mathbf{J}}(\boldsymbol{\ell})]_{1,2}&=&  \sum_{j=1}^M3 k_j \mu\tilde{x}_j\tilde{y}_j \times \left[ \frac{1}{\sigma_{d_j}^2 d_j^2}-\frac{1}{\sigma_{\phi_j}^2 d_{2,j}^4}\right.\nonumber \\ & &\left.+\bigg(\frac{\tilde{z}_j}{\sigma_{\theta_j} d_j^2 d_{2,j}}\bigg)^2\right],
 \end{eqnarray} 
\begin{equation}
 [\boldsymbol{\mathbf{J}}(\boldsymbol{\ell})]_{1,3}= \sum_{j=1}^M3 k_j \mu\frac{\tilde{x}_j\tilde{z}_j}{d_j^2}\left[\frac{1}{\sigma_{d_j}^2}-\frac{1}{\sigma_{\theta_j}^2d_j^2}\right],
 \end{equation}
 \begin{eqnarray}
 [\boldsymbol{\mathbf{J}}(\boldsymbol{\ell})]_{2,2}&=&  \sum_{j=1}^M 3 k_j \mu\left[ \bigg(\frac{\tilde{y}_j}{\sigma_{d_j}d_j}\biggl)^2+ \bigg(\frac{\tilde{x}_j}{\sigma_{\phi_j}d_{2,j}^2}\biggl)^2 \right.\nonumber\\
 & & \left. + \bigg(\frac{\tilde{y}_j\tilde{z}_j}{\sigma_{\theta_j} d_j^2 d_{2,j}}\bigg)^2 \right],
 \end{eqnarray}
 \begin{equation}
 [\boldsymbol{\mathbf{J}}(\boldsymbol{\ell})]_{2,3}=   \sum_{j=1}^M 3 k_j \mu \frac{\tilde{y}_j\tilde{z}_j}{d_j^2}\left[\frac{1}{\sigma_{d_j}^2}-\frac{1}{\sigma_{\theta_j}^2d_j^2}\right],
 \end{equation}
 \begin{equation}
 [\boldsymbol{\mathbf{J}}(\boldsymbol{\ell})]_{3,3}=  \sum_{j=1}^M 3 k_j \mu \frac{1}{d_j^2}\left[\bigg(\frac{\tilde{z}_j}{\sigma_{d_j}}\bigg)+\bigg(\frac{d_{2,j}}{\sigma_{\theta_j}d_j}\bigg)^2\right].
 \end{equation}
Note that $[\boldsymbol{\mathbf{J}}(\boldsymbol{\ell})]_{k,q} = [\boldsymbol{\mathbf{J}}(\boldsymbol{\ell})]_{q,k}$, $ d_{2,j} = \sqrt{\tilde{x}_j^2+\tilde{y}_j^2}$, $ k_j = \frac{P_{t_j} A_j \cos\theta}{2\pi (1-\cos\theta_0)}$, and $\mu = \exp(-c(\lambda))$. The terms in $k_j$ and $\mu$ consists of the transmit power from the $j$-th anchor $P_{t_j}$, aperture area of the receiver $A_j$, trajectory of the signal $\theta$, divergance angle $\theta_0$, and extinction co-efficient of the water $c(\lambda)$ respectively. Finally, the CRLB of the 3D localization is estimated as $ \text{CRLB}(\dot{\boldsymbol{\ell}}) = \text{CRLB}(\dot{x})+\text{CRLB}(\dot{y})+\text{CRLB}(\dot{z})$,
 where
 \begin{equation}\label{eq: crlb1}
 \text{CRLB}(\dot{x}) = \frac{[\boldsymbol{\mathbf{J}}(\boldsymbol{\ell})]_{2,2}[\boldsymbol{\mathbf{J}}(\boldsymbol{\ell})]_{3,3}-[\boldsymbol{\mathbf{J}}(\boldsymbol{\ell})]_{2,3}^2}{|[\boldsymbol{\mathbf{J}}(\boldsymbol{\ell})]|},
 \end{equation}
 \begin{equation}\label{eq: crlb2}
 \text{CRLB}(\dot{y}) = \frac{[\boldsymbol{\mathbf{J}}(\boldsymbol{\ell})]_{1,1}[\boldsymbol{\mathbf{J}}(\boldsymbol{\ell})]_{3,3}-[\boldsymbol{\mathbf{J}}(\boldsymbol{\ell})]_{1,3}^2}{|[\boldsymbol{\mathbf{J}}(\boldsymbol{\ell})]|},
 \end{equation}
 \begin{equation}\label{eq: crlb3}
 \text{CRLB}(\dot{z}) = \frac{[\boldsymbol{\mathbf{J}}(\boldsymbol{\ell})]_{1,1}[\boldsymbol{\mathbf{J}}(\boldsymbol{\ell})]_{2,2}-[\boldsymbol{\mathbf{J}}(\boldsymbol{\ell})]_{1,2}^2}{|[\boldsymbol{\mathbf{J}}(\boldsymbol{\ell})]|},
 \end{equation}
$(\dot{z},\dot{y},\dot{z})$  is the final estimate of the source location, and $|\cdot|$ represents the determinant operation. It is clear from above that the accuracy of the CRLB depends on that of  the ranging measurements, placement of anchors, and estimation of angles.

 \section{Analysis with Uncertainty in Anchor Positions}
Although it is a common practice to assume that the position of an anchor is perfectly known, this is not a realistic assumption, especially in the underwater environment where anchors can swing and drift due to surface waves and deep currents. Therefore, it is necessary to develop more practical localization techniques which account for the error in the anchor positions. The error in anchor position can be modeled as an additive term to the observation as $\hat{\text{\textramshorns}}_i = \hat{g}_i(\boldsymbol{\ell})+\eta_i$ where 
\begin{equation}
\label{eq:measurement}
\hat{g}_i(\boldsymbol{\ell})= \begin{cases}
\bar{d}_j = \sqrt{\bar{x}_j^2+\bar{y}_j^2+\bar{z}_j^2}+\eta_{\bar{d}_j} &, i=3j-2 \\
\bar{\phi}_j = \tan^{-1}\left(\frac{\bar{y}_j}{\bar{x}_j}\right)+\eta_{\bar{\phi}_j} &, i=3j-1 \\
\bar{\theta}_j=\cos^{-1}\left(\frac{\bar{z}_j}{\hat{g}_{3j-2}(\boldsymbol{\ell})}\right)+\eta_{\bar{\theta}_j} &,i=3j. \\
\end{cases} 
\end{equation}
The notations of $\bar{x}_j$, $\bar{y}_j$ and $\bar{z}_j$ are equal to $x-\hat{x}_j$, $y-\hat{y}_j$, and $z-\hat{z}_j$ respectively. Let the error for each coordinate of anchor $j$ to be $\Delta x_j = x_j - \hat{x}_j$, $\Delta y_j = y_j - \hat{y}_j$, and  $\Delta z_j = z_j - \hat{z}_j$ respectively. Then, the $x$, $y$, and $z$ coordinate errors for anchor $j$ are modeled as Gaussian random variables with zero mean and variances of  $\sigma_{\Delta x_j}^2$, $\sigma_{\Delta y_j}^2$, and $\sigma_{\Delta z_j}^2$, respectively \cite{Angjelichinoski2015}. In order to compute $\eta_{\bar{d}_j}$, $\eta_{\bar{\phi}_j}$, $\eta_{\bar{\theta}_j}$, we assume that the noise is small and therefore the following trigonometric approximations holds:
\begin{equation}
\tan({\phi}_j+\eta_{\bar{\phi}_j})\approx \frac{\sin(\phi_j)+\eta_{\bar{\phi}_j}\cos(\phi_j)}{\cos(\phi_j)-\eta_{\bar{\phi}_j}\sin(\phi_j)},
\end{equation}
\begin{equation}
\cos({\theta}_j+ \eta_{\bar{\theta}_j} )\approx \cos({\theta}_j)-\eta_{\bar{\theta}_j}\sin(\theta_j).
\end{equation}
By using the above approximations and after making some algebraic manipulations, we obtain 
\begin{equation}
\eta_{\bar{d}_j} \approx \frac{\tilde{x}_j\Delta x_j+ \tilde{y}_j\Delta y_j + \tilde{z}_j\Delta z_j}{d_j},
\end{equation}
\begin{equation}
\eta_{\bar{\phi}_j} \approx \frac{\cos (\phi_j)\Delta y_j - \sin (\phi_j)\Delta x_j}{\cos (\phi_j)(\tilde{x}_j)+\sin (\phi_j)(\tilde{y}_j)},
\end{equation}
\begin{equation}
\eta_{\bar{\theta}_j} \approx -\frac{\Delta z_j}{d_j \sin(\theta_j)}.
\end{equation} 
Now consider $\boldsymbol{\upsilon}_j = [\eta_{\bar{d}_j},\eta_{\bar{\phi}_j},\eta_{\bar{\theta}_j}]^T$ with co-variance $\boldsymbol{\mathbf{C}}_j = \mathds{E}[\boldsymbol{\upsilon}_j \boldsymbol{\upsilon}_j^T]$ where the elements of the co-varaince matrix are given as
\begin{equation}
[\mathbf{C}_j]_{1,1} = \frac{\sigma_{\Delta x_j}^2 \tilde{x}_j^2+\sigma_{\Delta y_j}^2 \tilde{y}_j^2+\sigma_{\Delta z_j}^2 \tilde{z}_j^2}{d_j^2}
\end{equation}
\begin{equation}
[\mathbf{C}_j]_{1,2} = \frac{\sin(\phi_j)\sigma_{\Delta x_j}^2\tilde{x}_j+\cos(\phi_j)\sigma_{\Delta y_j}^2\tilde{y}_j}{\bigg(\tilde{x}_j\cos(\phi_j)+\tilde{y}_j\sin(\phi_j)\bigg)d_j}
\end{equation}
\begin{equation}
[\mathbf{C}_j]_{1,3} = \frac{-\tilde{z}_j\sigma_{\Delta z_j}^2}{\sin(\theta_j)d_j^2},
\end{equation} 
 \begin{equation}
[\mathbf{C}_j]_{2,2} = \frac{\sin^2(\phi_j)\sigma_{\Delta x_j}^2+ \cos^2(\phi_j)\sigma_{\Delta y_j}^2}{\bigg(\tilde{x}_j\cos(\phi_j)+\tilde{y}_j\sin(\phi_j)\bigg)^2},
\end{equation} 
  \begin{equation}
[\mathbf{C}_j]_{2,3} = 0, \text{ and } 
[\mathbf{C}_j]_{3,3} = \bigg(\frac{\sigma_{\Delta z_j}}{\sin(\theta_j)d_j}\bigg)^2.
\end{equation}
Matrix ${\mathbf{C}}_j$ is symmetric, hence, $[\mathbf{C}_j]_{k,q}= [\mathbf{C}_j]_{q,k}$. Based on the co-variance matrix, the PDF of $\hat{\textbf{\textramshorns}}_j$ is written as
\begin{equation}\label{eq: pdf2}
 p(\hat{\textbf{\textramshorns}}_j|\boldsymbol{\ell})=\frac{\exp^{\bigg(-\frac{1}{2}(\hat{\textbf{\textramshorns}}_j - \boldsymbol{g}_j(\boldsymbol{\ell}))^T{\mathbf{C}}_j^{-1}(\hat{\textbf{\textramshorns}}_j - \boldsymbol{g}_j(\boldsymbol{\ell}))\bigg)}}{\sqrt{(2\pi)^{3}|{\mathbf{C}}_j|}} .
 \end{equation}
Following from the independence of measurements taken from different anchors, the conditional PDF of the ranging vector $\hat{\textbf{\textramshorns}}_j$ can be expressed as 
\begin{equation}\label{eq: pdf4}
 p(\tilde{\textbf{\textramshorns}}|\boldsymbol{\ell}) = \frac{\exp^{\bigg(-\frac{1}{2}\sum_{j=1}^M(\tilde{\textbf{\textramshorns}}_j - \boldsymbol{g}_j(\boldsymbol{\ell}))^T\boldsymbol{\mathbf{C}}_j^{-1}(\tilde{\textbf{\textramshorns}}_j - \boldsymbol{g}_j(\boldsymbol{\ell}))\bigg)}}{\sqrt{(2\pi)^{3M}\prod_{j=1}^M|\boldsymbol{\mathbf{C}}_j|}} .
 \end{equation}


%
%

 Accordingly, the elements of the FIM are derived from \eqref{eq: pdf4} as follows
\begin{equation}
 [\boldsymbol{\mathbf{J}}(\boldsymbol{\ell})]_{1,1}=\sum_{j=1}^M 3 k_j \mu \frac{\partial \boldsymbol{g}_j^T(\boldsymbol{\ell})}{\partial x} {\mathbf{C}}_j^{-1} \frac{\partial \boldsymbol{g}_j(\boldsymbol{\ell})}{\partial x},
\end{equation}
 \begin{equation}
 [\boldsymbol{\mathbf{J}}(\boldsymbol{\ell})]_{1,2}=[\boldsymbol{\mathbf{J}}(\boldsymbol{\ell})]_{2,1}=\sum_{j=1}^M 3 k_j \mu \frac{\partial \boldsymbol{g}_j^T(\boldsymbol{\ell})}{\partial y} {\mathbf{C}}_j^{-1} \frac{\partial \boldsymbol{g}_j(\boldsymbol{\ell})}{\partial x},
\end{equation}
\begin{equation}
 [\boldsymbol{\mathbf{J}}(\boldsymbol{\ell})]_{1,3}=[\boldsymbol{\mathbf{J}}(\boldsymbol{\ell})]_{3,1}=\sum_{j=1}^M 3 k_j \mu \frac{\partial \boldsymbol{g}_j^T(\boldsymbol{\ell})}{\partial z} {\mathbf{C}}_j^{-1} \frac{\partial \boldsymbol{g}_j(\boldsymbol{\ell})}{\partial x},
\end{equation}
 \begin{equation}
 [\boldsymbol{\mathbf{J}}(\boldsymbol{\ell})]_{2,2}=\sum_{j=1}^M 3 k_j \mu \frac{\partial \boldsymbol{g}_j^T(\boldsymbol{\ell})}{\partial y} {\mathbf{C}}_j^{-1} \frac{\partial \boldsymbol{g}_j(\boldsymbol{\ell})}{\partial y},
\end{equation}
 \begin{equation}
 [\boldsymbol{\mathbf{J}}(\boldsymbol{\ell})]_{2,3}=[\boldsymbol{\mathbf{J}}(\boldsymbol{\ell})]_{3,2}=\sum_{j=1}^M 3 k_j \mu \frac{\partial \boldsymbol{g}_j^T(\boldsymbol{\ell})}{\partial z} {\mathbf{C}}_j^{-1} \frac{\partial \boldsymbol{g}_j(\boldsymbol{\ell})}{\partial y},
\end{equation}
\begin{equation}
 [\boldsymbol{\mathbf{J}}(\boldsymbol{\ell})]_{3,3}=\sum_{j=1}^M 3 k_j \mu \frac{\partial \boldsymbol{g}_j^T(\boldsymbol{\ell})}{\partial z} {\mathbf{C}}_j^{-1} \frac{\partial \boldsymbol{g}_j(\boldsymbol{\ell})}{\partial z},
\end{equation}
where
\begin{equation}
\frac{\partial \boldsymbol{g}_j^T(\boldsymbol{\ell})}{\partial x} = \left[\frac{\tilde{x}_j}{d_j}~~~-\frac{\tilde{y}_j}{d_{2,j}^2}~~~\frac{\tilde{x}_j\tilde{z}_j}{d_j^2 d_{2,j}}\right],
\end{equation}
\begin{equation}
\frac{\partial \boldsymbol{g}_j^T(\boldsymbol{\ell})}{\partial y} = \left[\frac{\tilde{y}_j}{d_j}~~~-\frac{\tilde{x}_j}{d_{2,j}^2}~~~\frac{\tilde{y}_j\tilde{z}_j}{d_j^2 d_{2,j}}\right],
\end{equation}
and
\begin{equation}
\frac{\partial \boldsymbol{g}_j^T(\boldsymbol{\ell})}{\partial z} = \left[\frac{\tilde{z}_j}{d_j}~~~0~~~\frac{d_{2,j}}{d_j^2}\right].
\end{equation}
By substituting the above elements of the FIM in \eqref{eq: crlb1} to \eqref{eq: crlb3}, the CRLB with anchors positions uncertainty can be calculated.
\begin{figure}[t]
\captionsetup{justification=centering,font=small}
        \centering
\includegraphics[width=0.55\textwidth]{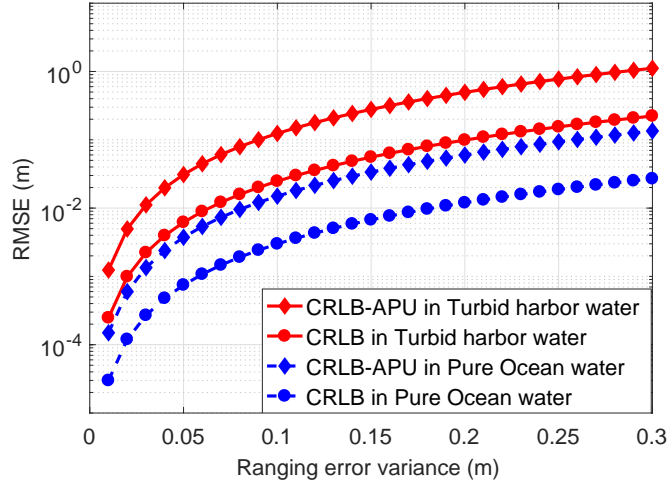}  
\caption{RMSE vs. range error\label{fig:opt_crlb}}
\end{figure}
\section{Numerical Results}
\begin{figure}[t]
 \centering
 \includegraphics[width=0.65\columnwidth]{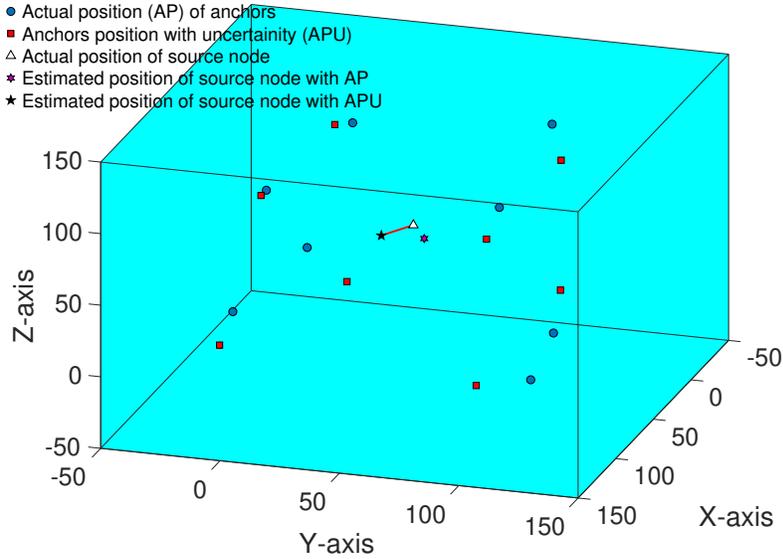}  
 \caption{3D scenario with anchor position uncertainity.\label{fig:scenario}}
 \setlength{\belowcaptionskip}{-10pt}  
 \end{figure}

\vspace{-0.0cm}
In this section, we perform numerous simulations to validate the performance of the derived analytical expressions in Section II and III. Fig. \ref{fig:scenario} shows a setting with 8 anchors nodes and a source node which need to be localized in $100~\text{m}^3$ cubic region where the blue circles, red square, and green triangle represents the actual position of anchor nodes, the drifted positions of the anchor nodes, and the actual position of the source node respectively. The purple hexagram and the black pentagram represent the estimated position of source node based on the actual positions and  drifted positions of anchors,  respectively. The standard deviation of the anchor positions error and ranging error is kept to 1.5 m and 1 m, respectively. Fig. \ref{fig:scenario} clearly demonstrates that the uncertainty in anchor positions results in low localization accuracy.


The results are averaged out over 1000 Monte Carlo simulations with $8$ random anchor nodes and a source node in  $100~\text{m}^3$ cubic region. $\text{RMSE}\triangleq \sqrt{\mathds{E}((\dot{\boldsymbol{\ell}}-{\boldsymbol{\ell}})^2)}$ is considered as the performance metric. The operating wavelength of the optical signals is in the blue region with a wavelength of 445 nm, and the divergence angle is equal to 30 degrees.

To show the effect of underwater optical channel parameters, we have considered two different types of water in Fig.~\ref{fig:opt_crlb}, i.e., pure ocean water and turbid harbor water. It is clear from Fig.~\ref{fig:opt_crlb} that for a given ranging error variance, the RMSE performance is better in pure ocean water due to low absorption and scattering loss as compared to the turbid harbor water.  Also, Fig.~\ref{fig:opt_crlb} shows that when there is anchor position uncertainty, the RMSE increases. Note that the range measurements depends on not only the underwater optical channel impediments but also the distance among the anchor nodes and the source node by using (14). It is worthy to note here that for the rest of simulations we consider pure ocean water.

 \begin{figure}[t]
 \centering
 \includegraphics[width=0.55\columnwidth]{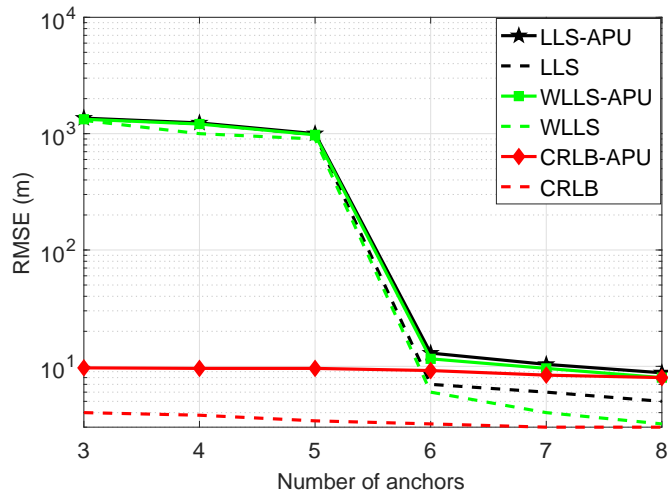}  
 \caption{RMSE versus number of anchors.\label{fig:rmse_anchors}}
 \vspace{-0.7 em} 
 \end{figure}
 \begin{figure}[t]
 \centering
 \includegraphics[width=0.55\columnwidth]{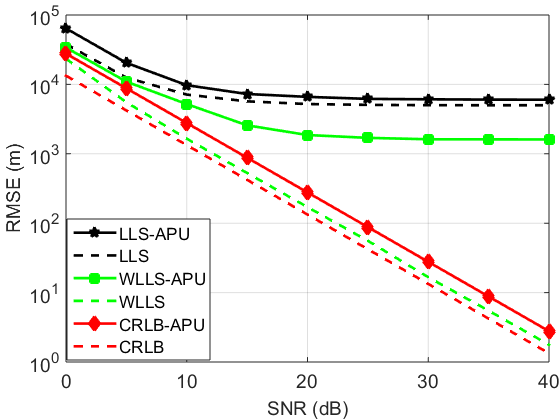}  
 \caption{RMSE versus ranging error variance with uncertainity in anchor positions, i.e., $\sigma_{\hat{x}}=\sigma_{\hat{y}}=\sigma_{\hat{z}} = 1.5$ m.\label{fig:rmse_ranging}}  
 \vspace{-0.5 em}
\end{figure}

Fig.~\ref{fig:rmse_anchors} shows the localization accuracy of the analytical results of CRLB, LLS, and WLLS concerning the number of anchors. The standard deviations for ToA and AoA measurements are equal to 2 m and 2 degrees, respectively. Fig.~\ref{fig:rmse_anchors} depicts that increasing the number of anchors improves the localization accuracy, and the LLS and WLLS methods achieve the CRLB when the number of anchors is equal to 8.  Fig.~\ref{fig:rmse_anchors} also shows that the CRLB with APU has a large localization error but is more practical as compared to the case when there is no uncertainty in anchors positions.

Next, we examine the performance of the LLS and WLLS in comparison with the derived CRLB for perfect and erroneous anchor positions, respectively. The uncertainty in each coordinate of the anchor position is modeled as a Gaussian random variable with zero mean and standard deviation of 1.5 meters. It is clear from Fig.~\ref{fig:rmse_ranging} that the WLLS solution achieves the CRLB with error-free anchor positions. However, in the presence of APUs, both LLS and WLLS have a large localization error. Notice that the SNR for Fig.~\ref{fig:rmse_ranging} is defined as the mean squared distance over the noise variance.

\section{Conclusion}\label{conc}
In this paper, we have derived a closed-form expression for the lower bound of error variance for 3D UOWSN localization in the presence of APU. The localization error is analyzed for both cases with and without APU by using ToA and AoA based ranging. The analytical findings have been validated by the numerical results where the error in anchor positions yields a low localization accuracy. Therefore, it is important for any 3D UOWSNs localization technique to account for anchor positions error in addition to the underwater optical channel parameters and noisy range measurements.


\end{document}